# A FASTER ROUTING SCHEME FOR STATIONARY WIRELESS SENSOR NETWORKS – A HYBRID APPROACH

Jasmine Norman[1] , J.Paulraj Joseph[2] , P.Prapoorna Roja[3]

[1]Measi Institute of Information Technology, Chennai – 14, India

**ddjnorman@yahoo.com**

[2] Manonmaniam Sundaranar University, Tirunelveli-12, India

[3] Jerusalem College of Engineering, Chennai – 100, India

## Abstract

*A wireless sensor network consists of light-weight, low power, small size sensor nodes. Routing in wireless sensor networks is a demanding task. This demand has led to a number of routing protocols which efficiently utilize the limited resources available at the sensor nodes. Most of these protocols are either based on single hop routing or multi hop routing and typically find the minimum energy path without addressing other issues such as time delay in delivering a packet, load balancing, and redundancy of data. Response time is very critical in environment monitoring sensor networks where typically the sensors are stationary and transmit data to a base station or a sink node.  In this paper a faster load balancing routing protocol based on location with a hybrid approach is proposed.*

## 1. INTRODUCTION

A wireless sensor network (WSN) consists of spatially distributed autonomous sensors to cooperatively monitor physical or environmental conditions, such as temperature, sound, vibration, pressure, motion or pollutants. The emergence of wireless sensor networks has enabled new classes of applications that benefit a large number of fields. In [1] Joseph Polastre et al have identified the need for using WSN for habitat and environmental monitoring. The challenges in the hierarchy of: detecting the relevant quantities, monitoring and collecting the data, assessing and evaluating the information, formulating meaningful user displays, and performing decision-making and alarm functions are enormous as suggested by F.L.Lewis [2]. Unique characteristics of a WSN include limited power, ability to withstand harsh environmental conditions, ability to cope with node failures, mobility of nodes, dynamic network topology, communication failures, heterogeneity of nodes, large scale of deployment and unattended operation.

Many to one communication paradigm is widely used in regard to sensor networks since sensor nodes send their data to a common sink for processing. This many-to-one paradigm also results in non-uniform energy drainage in the network. Sensor networks can be divided in two classes as event driven and continuous dissemination networks according to the periodicity of communication. In event-driven networks, data is sent whenever an event occurs. In continuous dissemination networks, every node periodically sends data to the sink. Routing protocols are usually implemented to support one class of network in order to save energy.

The challenges of WSN have been studied by Yao K [3]. The key challenge in wireless sensor networks is maximizing network lifetime. The appropriate communication mode will significantly reduce energy consumption of communication and prolong networks lifetime. Therefore, many researchers are currently focusing on the design of power-aware protocols for wireless sensor networks. Regardless of communication protocol, researchers must choose communication mode: single hop or multi hop. Using a single hop communication mode, each sensor sends its data directly to the base station. In multi hop mode, each node sends its data destined ultimately for the base station through intermediate nodes. Now multi hop communication enjoys more researches' favor. In [13], Bhardwaj et al. have studied a multi hop sensor network; they minimized the energy spent on sending a packet by using optimum number of relay nodes. Bandyopadhyay et al. [14] have studied a multi hop clustered wireless sensor network. [8,11] give a comparative study of multihop routing protocols. The reason many researches choose multi hop mode lies in that it is expected to consume less power than the single hop communication, but that is not always correct. In most wireless sensor networks nodes are static, using multi hop mode, nodes closest to the base station have a highest load of relaying packets as compared to other nodes, just as the nodes located farthest away from the base station have the highest energy burden due to long range communication in single hop mode. These key nodes will quickly drain the battery and result in invalidation of the whole system, although other nodes have enough energy. It is evident from [20] that multi hop communication is not always the best mode.

Monitoring the environment is one of the main applications of wireless sensor networks. Given that these networks are densely populated and that local variations in the environmental variables are small, a large amount of redundant data is generated by sensor nodes. The time delay in receiving information in the base station end is very critical to the better functioning of the network. When an event occurs all the sensors in the region will sense and start transmission of data. The base station will receive the same data from a number of sensors. Also transmission energy is greater than the processing energy. Thus energy is wasted to transmit redundant information to the base station. So there is a trade off between time and reliability and in WSN reliability can be compromised as redundancy.

Most of the protocols suggested are based on energy conservation. The redundancy of information from the sensors is not taken into consideration. Again most of the protocols assume multi hop paths. Single hop networks also proved to be energy efficient. Thus there is a need to approach the problem in a balanced way. This paper argues the energy savings due to single hop and presents an efficient algorithm to use the hybrid structure.

## 2. RELATED WORK

Sensor networks introduce new challenges that need to be dealt with as a result of their special characteristics. Their new requirements need optimized solutions at all layers of the protocol stack in an attempt to optimize the use of their scarce resources. In particular, the routing problem, has received a great deal of interest from the research community with a great number of proposals being made. Basically these protocols can be fit in one of two major categories: on-demand such as AODV [4] and DSR [5], and proactive such as DSDV [6] and OLSR [7]. The review and performance comparison of these protocols are in [8,9,10.11]. Directed diffusion [15] is a good candidate for robust multi hop multipath routing and delivery.

The common belief is that a multi-hop configuration with rather small per-hop distance is the only viable energy efficient option for wireless sensor networks.

Cellular networks, WiFi and many other single hop networks have used single hop structures not for energy considerations, but for other reasons: infrastructure constraints, simpler network management and the other benefits of a single hop structure enumerated in the previous section. There also exist designs developed for single hop sensor networks [19,21,22]. These work choose a single hop network not because of its energy efficiency, but because it is a less complicated network.

Location-based algorithms [16,17,18] rely on the use of nodes position information to find and forward data towards a destination in a specific network region. Position information is usually obtained from GPS (Global Positioning System) equipment. They usually enable the best route to be selected, reduce energy consumption and optimize the whole network. In [18] Ye Ming Luz et al have proposed location based energy efficient protocol. Na Wang et al in [9] have studied the performance of the geographic based protocols.

It is proved in [20] that the single hop cost less communication energy than multi-hop when amount of sensor nodes and communication radius is little and multi-hop mode is more effective unless amount of sensor nodes and communication radius is very small when value of propagation loss exponent becomes larger. When a realistic radio model is applied for a sensor network, it was discovered that with feasible transmission distances single-hop communications can be more efficient than multi-hop in the energy perspective.In [19] Lizhi Charlie Zhong et al discuss a single hop configuration, utilizing the asymmetry between lightweight sensor nodes and a more powerful "base station" and demonstrate that such a single hop configuration can actually have lower overall power consumption than a multi-hop counterpart.

Existing wireless sensor routing protocols commonly use minimum hop count as metric to find routes, and under two assumptions: a link which is good for route discovery messages is still good for data packets; secondly, the link quality is binary: either very good or very bad. Protocols such as DSR [5] and AODV [4] use broadcast messages to find the shortest paths, when the node receives the route reply, it will use this route to transmit data. Thus in additions to the data packets, a lot of control packets are generated adding to the traffic congestion, which will have an effect on the delivered time to the base station. This could prove to be very costly. This paper proposes a model based on location which takes into account load distribution, energy and redundancy as the main parameters, and maximizes the network lifetime with faster delivery time.

## 3. SYSTEM MODEL

The network is assumed to be static and fairly distributed with a single base station. All the nodes are assumed to know their location as well as destination's location. The model proposed is used for making a decision on which neighbour a sensor node should forward the data message to in case if it is not able to deliver the message straight to the base station. There are two phases in the proposed model.

### 3.1 Initial Configuration Phase

After deployment each sensor sends its location to the base station. The base station prepares the neighbour hood table of each sensor and forwards. The neighbour discovery process is location oriented. Each node must have a minimum specified number of neighbours.

The region is divided into vertical areas. If the location of the node is (x,y) all nodes within the range x+i ,y+j are its neighbours where i <= M (a specific integer) and j <= N , any number. Since all the nodes are assigned neighbours, all the sensors equally participate in the transmission.

## 3.2 Transmission / Receive Phase

When a node senses an event , it compares the residual energy (r) and the threshold energy (tr) levels. If r < tr, then the node has no more energy to take any transmission job and it goes off to sleep. If the node has sufficient energy, it checks the data buffer for an equivalent entry. If a match is not found, it measures the strength of the received signal. If it finds one, the packet will be discarded. The node calculates the feasibility of a single hop transmission taking into account SNR and location of the destination. If feasible, it sends the data directly to the destination. Otherwise the node computes the best neighbor and forwards data. The best neighbour is computed based on SNR, Residual energy level, use count and the location. The farthest location node in the vertical region towards the destination will be examined first.

The same process will be repeated till the packet reaches the destination.  After transmission, the nodes that were involved in the transmission send their residual energy to the base station so that they can update their routing table. Periodically the neighbourhood table is updated by the base station leaving the dead nodes. There are no control packets overhead in this model as against AODV[4] and DSR[5]. This model suppresses redundant data packets but there is no data aggregation as in Directed Diffusion [15] which will consume more energy. [18] proposes an energy efficient location oriented protocol but does not take care of load balancing and redundant data suppression.

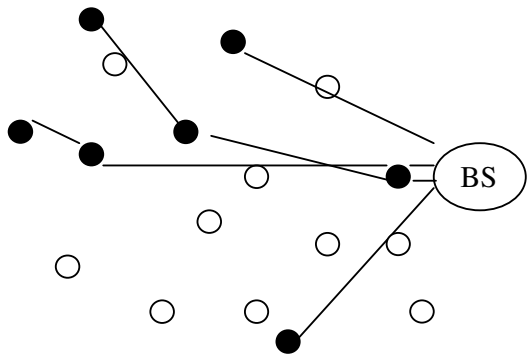

Figure 1: Path Selected for sample nodes

## 3.3 Neighbourhood Table Computation

Assume the sensors are fairly distributed. Each sensor node sends the location and the residual energy to the base station.  The base station computes the neighbours for each sensor by using the location. A sample format of the location table and neighbourhood table are given below. Each sensor has a unique ID. In the neighbour table the first column represents the sensor ID and the rest are the best neighbours computed based on the location towards the destination. 0 in the best neighbour column indicates that the node is positioned close to the base station. The neighbours are arranged in the order of closer proximity to the base station.

<u>Location Table</u>

0 , 58 , 258
1 , 160 , 275
2 , 163 , 192
3 , 216 , 202
4 , 205 , 166
5 , 167 , 227

**Neighbour Table**

| 0 | 0  | 0  | 0  |
|---|----|----|----|
| 1 | 67 | 66 | 50 |
| 2 | 5  | 69 | 43 |
| 3 | 5  | 70 | 33 |
| 4 | 3  | 39 | 37 |
| 5 | 0  | 0  | 0  |

### 3.4 Best Neighbour Algorithm

This is based on residual energy level of the neighbour, signal strength, usage count and the location.

1. Pick  the first neighbour with the least use count
2. If  (residual energy level > threshold )
3. If  (SNR  is high)  Forward the data
   Else  Repeat the process eliminating the current node explored.

## 4. ADVANTAGES OF THE SYSTEM

**Faster Delivery Time**  - If the SNR is high , the delay in multi hop can be avoided. No control packets are used.

**Energy Efficient  -**  Utilizing the best of single hop and multi hop

**Collision Management -** When a channel is busy the source wishes to transmit chooses the next best neighbour and transfers the data. If two nodes choose the same neighbour, depending on signal strength (SNR) the node picks one and discards the other.

**Congestion Control - O**nly mutually exclusive paths are explored. An already visited node will not be included in the path. It speeds up the entire process avoiding congestion in one route.

**Load Balancing -** Whichever node that senses an event can initiate the transmission and it can use only its neighbours with least use count to forward the data. Thus without draining the already used nodes, network life time can be extended.

**Fault Tolerant -** After a specific period of transmission the base station computes the neighbourhood table for each sensor upon receiving the energy level from the nodes. The dead nodes will be eliminated from the network.

**QOS** – It is guaranteed to deliver data to the destination quickly.

## 5. PERFORMANCE ANALYSIS

We simulate this protocol on GloMoSim, [23, 24] a scalable discrete-event simulator developed by UCLA. This software provides a high fidelity simulation for wireless communication with detailed propagation, radio and MAC layers. We compare the routing protocol named as HYB with two popular sensor networks routing protocols – AODV and DSR

### 5.1 Simulation Model

The GloMoSim library [24] is used for protocol development in sensor networks. The library is a scalable simulation environment for wireless network systems using the parallel discrete event simulation language PARSEC. The distributed coordination function (DCF) of IEEE 802.11 is used as the MAC layer in our experiments. It uses Request-To-Send (RTS) and Clear-To-Send (CTS) control packets to provide virtual carrier sensing for unicast data packets to overcome the well-known hidden terminal problem.

There are some initial values used in the simulation. Intel Research Berkeley Sensor Network Data and WiFi CMU data from Select Lab [25] are used to get the positions for the nodes. The experiment is repeated for varying number of nodes. CBR traffic is assumed in the model. The new protocol is written in Parsec and hooked to GloMoSim. All the three protocols are simulated in GloMoSim to enable comparisons among them. When a packet is generated, the corresponding routing algorithm is invoked.

**Table 1. Assumed Parameters**

| Parameters | Value |
|---|---|
| Transmission range | 250 m |
| Simulation Time | 5M |
| Topology Size | 2000m x 2000m |
| Number of sensors | 25, 50.75 |
| Number of sinks | 1 |
| Mobility | None |
| Traffic type | Constant bit rate |
| Packet rate | 8 packets/s |
| Packet size | 512 bytes |
| Radio Type | Standard |
| Packet Reception | SNR |
| Radio range | 350m |
| MAC layer | IEEE 802.11 |
| Bandwidth | 2Mb/s |
| Node Placement | Node File |
| Initial energy in batteries | 10 Joules |
| Signal Strength Threshold | -80 dbm |
| Energy Threshold | 0.001mJ |

## 5.2 Performance Metrics

For the evaluation of protocols the following metrics have been chosen. Each metric is evaluated as a function of the topology size, the number of nodes deployed, and the data load of the network.

*Execution Time* : It is the total time taken by the various protocols for the given CBR traffic to complete within the simulation time. This does not guarantee the reliability of data packets generated. The faster protocol may have less execution time but it may not guarantee the delivery of all the generated packets.

*Hop Count* : The number of hops used by the protocol to reach the destination.

*Collision* : The number of collisions occurred while delivering the packet. This is an indication of congestion in the traffic.

*Number of signals Transmitted* : The total number of signals used in the transmission. This is an indication of the energy conserved by the protocol.

## 5.3 Simulation Results

Figure 2 shows the execution time of three protocols for different sets of nodes and traffic. The execution time increases as the traffic increases. Due to control packets overhead in route discovery and maintenance AODV and DSR have high execution time as against the proposed protocol. But the proposed protocol does not guarantee the delivery of all the packets generated. This is based on the assumption that at a particular time the nodes sense the same data and transmit. So it is not necessary for the base station to receive all the redundant information. The data get suppressed in the intermediate nodes due to redundancy and bad link quality. Also if there is congestion the data packet will simply be discarded after waiting for a specific time 't'. The nodes once rejected the packet never try to retransmit after a specified time as in AODV.

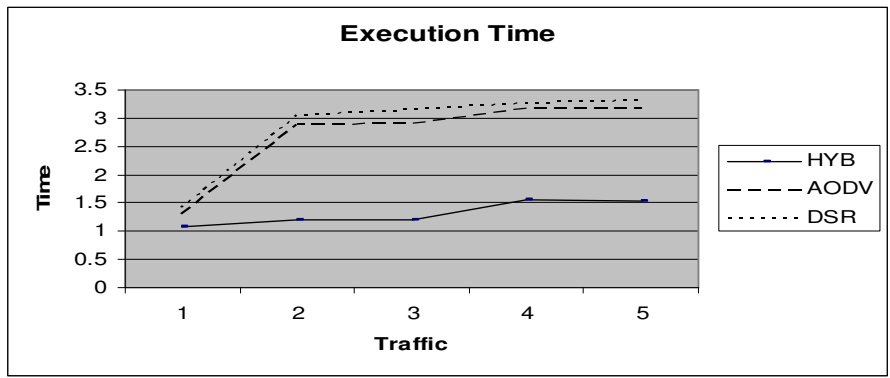

Figure 2. Execution Time

Figure 3 shows the average hop count where the proposed model has the lowest and AODV has the highest. If the link quality is good then direct transmission is possible. In this case the hop count will be equal to 0. The average hop count will always be less than or equal to the other models as it chooses the *best* neighbour at each cycle.

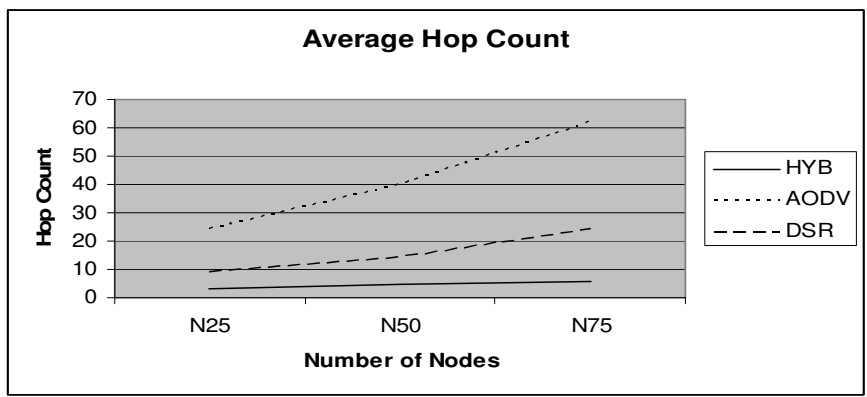

Figure 3. Average Hopcount

Figure 4 is an indication of congestion control in the various protocols. The more number of collisions indicate high traffic in a particular region. The redundancy suppression and SNR factor manage the congestion efficiently than AODV and DSR. Also retransmission is not suggested in the proposed model to manage the traffic efficiently.

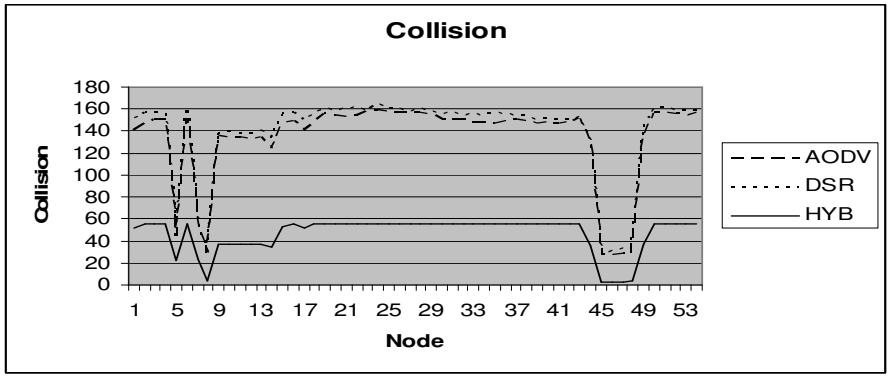

Figure 4. Number of Collisions

For maximizing the network lifetime energy conservation is important. Figure 5 shows the total number of signals transmitted for a given traffic The proposed model has transmitted less number of signals and therefore consumed less energy compared to the other protocols.

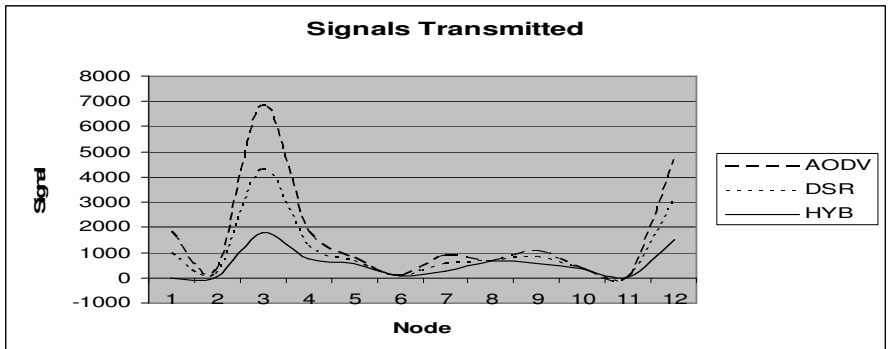

Figure 5. Transmission of Signals

# 6. CONCLUSION

Latency time is critical to the functioning of the environmental monitoring sensor networks. Also energy resource limitations are of priority concern in sensor networks. Distributing the load to the nodes significantly impacts the network life time. In this paper a faster energy efficient load balancing protocol for routing is proposed. It does fairly well compared to popular protocols DSR and AODV when simulated in GloMoSim environment. The proposed model has less time delay, more energy and better distributed work mode. This can be very effective for detecting environmental changes where sensors have a fixed location. The limitation in this protocol is that it does not guarantee the delivery of all the packets generated as retransmission is not supported. So there is a very small probability of missing relevant data. Our next step is to consider a faster mobile sensor network based on location and data aggregation.

# 7. REFERENCES

[1]     J. Polastre, R. Szewcyk, A. Mainwaring, D. Culler, J. Anderson, (2004) , "Analysis of Wireless
        Sensor Networks for Habitat Monitoring in Wireless Sensor Networks" , Kluwer Academic
        Publishers (NY),  pp. 399-423.

[2]     F. L. LEWIS (2004) "Wireless Sensor Networks" , Smart Environments: Technologies,
        Protocols, and     Applications , ed. D.J. Cook and S.K. Das, John Wiley

[3]     YAO Kung , (2006) ,  "Sensor Networking: Concepts, Applications, and Challenges", ACTA
        Automatica Sinica , Vol. 32, No. 6

[4]     C. E. Perkins, E. M. Royer, and S. R. Das,  "Ad Hoc On-Demand Distance Vector (AODV)
        Routing" , IETF Mobile Ad Hoc Networks Working Group, IETF RFC 3561

[5]     D. B. Johnson, D. A. Maltz, and Y-C Hu., (2003),  " The Dynamic Source Routing Protocol for
        Mobile Ad  Hoc Networks (DSR)" , IETF Mobile Ad Hoc Networks Working Group, Internet
        Draft

[6]     C. E. Perkins and P. Bhagwat, (1994) ,  "Highly dynamic destination-sequenced distance-vector
        routing (DSDV) for mobile computers" . In Proceedings of the ACM Special Interest Group on
        Data Communications  (SIGCOMM),  pp 234-244

[7]     T. Clausen, Ed., P. Jacquet, " Optimized Link State Routing Protocol (OLSR) " , Network
        Working  Group, Request for Comments: 3626

[8]      Bharat Kumar Addagada, Vineeth Kisara and Kiran Desai , (2009) , "A Survey: Routing
        Metrics  for Wireless Mesh Networks"

[9]     Na Wang  and Chorng Hwa Chang (2009) "Performance analysis of probabilistic multi-path
        geographic routing in wireless sensor networks" ,   International Journal of Communication
        Networks and Distributed Systems , Vol 2 , pp 16 – 39

[10]    A. H. Azni, Madihah Mohd Saudi, Azreen Azman, and Ariff Syah Johari D (2009) ,
        "Performance Analysis of Routing Protocol for WSN Using Data Centric Approach" , World
        Academy of Science, Engineering and Technology 53

[11]    Jamal N. Al-Karaki Ahmed E. Kamal , (2004) , "Routing Techniques in Wireless Sensor
        Networks: A Survey"

[12]     S. Das, R. Castaneda, and J. Yan, (2000) "Simulation-Based Performance Evaluation of Routing
        Protocols for  Mobile Ad Hoc Networks," Mobile Networks and Applications, Vol. 5, No.
        3, pp 179-189

[13]     Bhardwaj M, Garnett T and Chandrakasan A P (2001),  " Upper bounds on lifetime of sensor
        networks" , IEEE International Conference on Communications (Helsinki) pp 785-790

[14]    Bandyopadhyay S and Coyle E (2003) " An energy efficient hierarchical clustering algorithm for
        wireless  sensor networks" , IEEE Infocom 1713-23

[15]     C. Intanagonwiwat, R. Govindan, and  Estrin, (2000)  "Directed diffusion: A scalable and
        robust   communication paradigm for sensor networks," in Proc. of ACM MobiCom'00,
        Boston, MA,      USA,  pp. 56–67

[16]    Author:Zheng Kai, Tong Libiao, Lu Wenjun, (2009) ,  "Location-Based Routing Algorithms for
        Wireless Sensor Network " , ZTE Communications

[17]    Nicklas Beijar, "Zone Routing Protocol (ZRP)" , CiteSeer

[18]    Khalid Kaabneh, Azmi Halasa and Hussein Al-Bahadili , (2009) , "An Effective Location-Based
        Power Conservation Scheme for Mobile Ad Hoc Networks" , American Journal of Applied
        Sciences 6 (9) , pp  1708-1713

[19]     Lizhi Charlie Zhong , Jan M. Rabaey, Adam Wolisz , (2005) "Does Proper Coding Make Single
        Hop Wireless Sensor Networks Reality: The Power Consumption Perspective " , Proc. Of IEEE
        Wireless  Comm. and Networking Conf

[20]    J F Shi, X X Zhong and S Chen (2006) "Study on Communication Mode of Wireless Sensor
        Networks Based on Effective Result" , Journal of Physics: Conference Series 48 1317–1321

[21]     S. Saha and P. Bajcsy, (2003) "System design issues in single-hop wireless sensor  networks",
        *Proc. of 2nd  IASTED ICCIIT ,* Scottsdale, Arizona

[22]    M. Singh and V.K. Prasanna, (2003) "Optimal energy-balanced algorithm for selection in a
        single-hop sensor network", *IEEE international workshop on SNPA ICC*

[23]    M. Takai, L. Bajaj, R, Ahuja, R. Bagrodia and M. Gerla. (1999) ,  "GloMoSim: A Scalable
        Network  Simulation Environment", Technical report 990027, UCLA

[24]     Jorge Nuevo , "A Comprehensible GloMoSim Tutorial "

[25]    http://www.select.cs.cmu.edu/data/index.html